\documentclass[aps,prd,twocolumn,showpacs,amsmath]{revtex4}
\usepackage[dvips]{color,graphicx}
\usepackage{amsfonts}
\usepackage{amssymb}
\usepackage{latexsym}
\usepackage[latin1]{inputenc}
\usepackage[catalan]{babel}

\newcommand{\dalm}{\kern1pt\vbox{\hrule height 0.9pt\hbox{\vrule width
0.9pt\hskip 2.5pt\vbox{\vskip 5.5pt}\hskip 3pt\vrule width 0.3pt}\hrule height
0.3pt}\kern1pt}

\newcommand{\lw}[1]{\smash{\lower2.ex\hbox{#1}}}

\begin{document}


\title{Rigid motions and generalized Newtonian gravitation.\\ {\it Lost in Translation}}

\author{Xavier Ja\'en}
\affiliation{Departament de F\'isica i Enginyeria Nuclear, Universitat Polit\`ecnica de Catalunya, Spain}
 \email{xavier.jaen@upc.edu}

\author{Alfred Molina}
\affiliation{Departament de F\'isica Fonamental, Universitat de Barcelona, Spain}
\email{alfred.molina@ub.edu}

\date{\today}

\begin{abstract}
We try to lay down the foundations of a Newtonian theory where inertia and gravitational fields appear in a unified way aiming to reach a better understanding of the general relativistic theory. We also formulate a kind of equivalence principle for this generalized Newtonian theory. Finally we find the non-relativistic limit of the Einstein's equations for the space--time metric derived from the Newtonian theory.
\end{abstract}


\maketitle

\section{\label{Sec01}Introduction} Since Charles-Augustin de Coulomb in 1785 introduced the law of force for rest electrically charged particles till 1905 when Albert Einstein introduced the theory of special relativity, first electrostatics and later electromagnetism never stoped to evolve. This fact contrasts with the laws of gravity introduced by Newton in 1687 that remained unmodified untill ``impelled'' by the birth of special relativity, lead to the theory of general relativity in 1916. In this sudden evolution of Newtonian gravitation something was lost along the way. 

The concept of rigid motion in non-relativistic mechanics disappears swallowed by the Principle of general covariance\cite{Stachel}.
Furthermore the principle of equivalence doesn't establish clearly the kinship between inertial and gravitational fields. Also the principle of covariance has no physical meaning at all \cite{Krestchmann} and also conceals the fact that the General Theory of Relativity has no dynamic invariance group.

In  section II  we describe the notions of rigid motion and non-inertial reference frame, and write some field equations for the velocity field of the rigid motion. 

In section III we write the usual equations for the electromagnetic field and the Lorentz force using a suitable unit system and a gauge where only the vector potential is needed.  We also find a non-relativistic limit for the electromagnetic field equations and compare them with the usual Newtonian gravitational field in an inertial frame. 

In the section IV we propose a non-relativistic equivalence principle and introduce two generalizations of the Newtonian gravitation theory where, instead of a scalar potential, we have a velocity field which plays the same role as the velocity field used to describe a rigid motion. The second of these generalizations includes non-inertial reference frames and gravitational fields in the same formulation. 

In  section V we obtain the Lagrangian formulation of the theory where the role of the new principle of equivalence can be clearly expressed. 

Finally we write the space-time metric whose non-relativistic limit leads to the Newtonian gravitation. Of course this is not the first attempt to generalize the Newtonian theory   \cite{Cartan, Trautman, Havas, Kunzle, Ehlers}. This paper is strongly inspired by the first part of the article {\it Rigid motion invariance of Newtonian and Einstein's theories of General Relativity} by Ll. Bel\cite{LB}.

\section{\label{Sec02}Inertial observers, non-inertial observers and rigid motion}
In a Galilean frame of reference, according to the law of inertia, the  equation of motion for a free particle is:
\begin{equation}\label{eq01}
\frac{d^2 \vec{x}(t)}{dt^2} = 0
\end{equation}		

As it is well known, rigid motions are defined as those motions in which the Euclidean distances between space points remain constant. A non-inertial observer is defined by a clock (absolute time) and a rigid motion (three orthogonal axis moving rigidly). According to  Chasles theorem,  the most general motion of this rigid non-inertial frame can be decomposed in a rotation and a translation with respect to an inertial observer.

The same free particle described by \eqref{eq01} in an inertial frame can be described in a non-inertial frame with the origin  at $\vec{X}(t)$ and with a set of orthogonal axis rotating with an angular velocity  $\vec\Omega(t)$ with respect to the inertial frame.
The space points $\vec{x}$ can be written in the non-inertial system as $\vec{x}=\vec{X}+\vec{y}$ where $\vec{y}$ is a vector with origin at $\vec X$. The acceleration in the non-inertial frame is (Annex I)\eqref{eqA3}
\begin{equation}\label{eq02}
\vec b\equiv\frac{d^2 \vec{y} (t)}{dt^2 } =  -\vec A - \frac{{d\vec \Omega }}
{{dt}}\times \vec y -\vec \Omega  \times (\vec \Omega  \times \vec y) - 2\vec \Omega  \times \vec w 
\end{equation}	
where
\begin{equation}\label{eq03}
\vec w =\frac{d\vec y}{dt};\quad \vec A =\frac{d^2\vec X}{dt^2}
\end{equation}	
$\vec w$ is the particle velocity in the non-inertial frame. The particle behaves as if it where subjected to some ``inertial force fields'', $\vec{g}_I$ and $\vec{\beta}_I$  defined by
\begin{equation}\label{eq04}
\vec{g}_I\equiv-\left(\vec A + \frac{{d\vec \Omega }}{{dt}}\times \vec y +\vec \Omega  \times (\vec \Omega  \times \vec y)\right);\quad
\vec{\beta}_I\equiv 2\vec \Omega 
\end{equation}
Using these fields the equation of motion of a free particle in a non-inertial reference frame can be written as
\begin{equation}\label{eq05}
\frac{d^2 \vec{y}(t)}{dt^2 }=\vec{g}_I+\frac{d\vec y}{dt}\times\vec{\beta}_I
\end{equation}
That is, the equation of motion of a free particle can be written as (\ref{eq01}) in an inertial frame and as the equation (\ref{eq05}) in an non-inertial reference system. Alternatively we can write (\ref{eq05}) as
\begin{equation}\label{eq05b}
\frac{d^2 y^i}{ds^2 }=g_I^i+\eta^i_{jk} \frac{d y^j}{ds}\beta_I^k;\quad \frac{d^2 t}{ds^2}=0
\end{equation}
where $\eta^i_{jk}$ is the unit antisymmetric tensor of rank three.
From \eqref{eq05b} we can interpret the inertial fields as a Newtonian affine connection whose connection symbols vanish except $\Gamma^i_{00}=-{g}_I^i$ and $\Gamma^i_{j0}=-\eta^i_{jk}\Omega^k$, and therefore 
\begin{equation}\label{eq05c}
\frac{d^2 y^\mu}{ds^2 }+\Gamma^{\mu}_{\nu\rho}\frac{d y^\nu}{ds}\frac{d y^\rho}{ds }=0
\end{equation}
where $y^0\equiv t$. It can be easily seen that the class of these symmetric connections with these properties and $\Gamma^k_{j0} \delta_{ki}+\Gamma^k_{i0} \delta_{kj}=0$ are invariant by the rigid motions group.

The velocity field $ \vec v_0 (\vec x,t)$ for the rigid motions can be written as
\begin{equation}\label{eq06}
\vec v_0 (\vec x,t) = \frac{{d\vec X(t)}}
{{dt}} + \vec \Omega (t) \times ( \vec x-\vec X(t) )
\end{equation}	
where $\vec X(t)$ and  $\vec \Omega (t)$ are two vector fields, which are arbitrary functions of time.

This velocity field (\ref{eq06}) is usually found in the literature, but to our knowledge nobody has used it  as a ``vector potential'' from which the ``inertial force fields'', $\vec{g}_I$ and $\vec{\beta}_I$ can be derived as 
\begin{equation}\label{eq08}
\vec{g}_I=\vec \nabla \left( {\frac{{\vec{v}_0^2 }}
{2}} \right) -\frac{{\partial \vec v_0 }}{{\partial t}};\quad\vec \beta _I  =   \vec \nabla  \times \vec v_0
\end{equation}
the ``inertial force fields'' verify the following field equations
\begin{equation}\label{eq09}
\left. \begin{gathered}
  \vec \nabla  \times \vec g_I  =  - \frac{{\partial \vec \beta _I }}
{{\partial t}} ;\quad \vec \nabla \cdot\vec \beta _I  = 0 \hfill \\
  \vec \nabla \cdot\vec g_I  = \frac{1}
{2}\beta _I^2;\qquad \vec \nabla  \times \vec \beta _I  = 0 \hfill \\ 
\end{gathered}  \right\}.
\end{equation}
The first of these equations is formally identical to the electromagnetic Faraday  law, the second equation coincides with the Amp\`ere's law (no magnetic poles) and the third one is a nonlinear modification of the Coulomb's law without electrical charges.
 
In General Relativistic mechanics we have no way to characterize a class of physically relevant observers. This is why General Relativistic mechanics lacks of a dynamical symmetry group, contrarily to what happens in Newtonian mechanics. And in connection with this, a satisfactory way of inplementing the notion of kinematic rigidity is not known \cite{Born}, \cite {HN}.

\section{\label{Sec03} Electromagnetic interaction and non-relativistic gravitational interaction}
The purpose of this section is to analize the similarities and differences between the electromagnetic interaction and its limit $ c \rightarrow\infty $ on one hand and the Newtonian gravitational interaction on the other.
We will compare the corresponding electromagnetic field equations with those fullfilled for the non-inertial fields, equations \eqref{eq08} and \eqref{eq09}. 

The most common version of the basic equations for the electromagnetic interaction uses the rationalized MKS system of units, in which, besides the usual mechanical units, a new unit is added, the ampere. Here we will write the electromagnetic equations in a general unspecified system of units.  It is known that in the most general form of writing the electromagnetic equations we can introduce four constants $ k_i,\; i=1\ldots 4 $. \cite{Jackson}\cite{LlM}

\noindent 
\textbf{1)The Lorentz force}: the trajectory $\vec x $ of a particle of charge $ q $ and mass $ m $ inside an electromagnetic field, $\vec E $ and $\vec B $, (provided that the particle moves with a small velocity compared to the speed of light in order to avoid the relativistic linear momentum in the left hand side), is a solution of the equation of motion: 
 \begin{equation}\label{eq12}
m\frac{{d^2 \vec x}}
{{dt^2 }} = q\;\left( {\vec E + \;k_3 \frac{{d\vec x}}
{{dt}} \times \vec B} \right)
\end{equation}
Given the fields $ \vec E $ and $ \vec B $, equation \eqref{eq12} belongs to the non-relativistic mechanics. It is so as it is usually used. But, for \eqref{eq12} to be a genuinely non-relativistic equation, the fields $ \vec E $ and $ \vec B $, should also to be derived from non-relativistic equations and to transform in accordance with the Galileo transformations.

\noindent\textbf{2) The Maxwell equations}: the electromagnetic fields $ \vec E $ and $ \vec B $ are a solution of
\begin{equation}\label{eq13}
\left. \begin{gathered}
  \vec \nabla  \times \vec E =  - k_3 \frac{{\partial \vec B}}
{{\partial t}} ;\quad \vec \nabla \cdot\vec B = 0 \hfill \\
  \vec \nabla \cdot\vec E = 4\pi k_1 \rho  ;\quad \vec \nabla  \times \vec B = 4\pi k_4 k_2 \vec J + \frac{{k_4 k_2 }}
{{k_1 }}\frac{{\partial \vec E}}
{{\partial t}} \hfill \\ 
\end{gathered}  \right\}
\end{equation}
The constants  $k_i$ can be freely chosen with the following dimensional restrictions  
\begin{equation}\label{eq14}
  \left[ {k_1 } \right] = \left[ {k_2 } \right]L^2 T^{ - 2} ;\quad \left[ {k_3 k_4 } \right] = 1 ;\quad \frac{{\left[ E \right]}}
{{\left[ B \right]}} = \left[ {k_3 } \right]LT^{ - 1} ;
\end{equation}
and 
\begin{equation}\label{eq14b}\frac{{k_1 }}
{{k_2 k_3 k_4 }} = c^2
\end{equation}
where $c$ is the vacuum propagation speed for electromagnetic waves and 
$\rho $ and $ \vec J $ are respectively the charge and current densities.

Depending on the  application different unit systems have been chosen. These are described by different values $ k_i $, as shown in the table below: \cite{Jackson}\cite{LlM} 
\begin{center}
\begin{tabular}{| l | c | c | c | c |}
\hline
  &$ k_1 $& $ k_2 $& $ k_3 $& $ k_4 $\\
\hline
UES	&1  & $c^{ - 2}$ & $1$ & $1$\\
\hline
UEM	& $c^{ 2}$ & $1$ &	$1$ & $1$\\
\hline
GAUSS & $1$ & $c^{ - 2}$ & $c^{ - 1}$ & $c$\\
\hline
H-L &	$1/(4\pi)$ & $1/(4\pi c^2)$ & $c^{ - 1}$ & $c$\\
\hline
MKS-rac.& $1/(4\pi \varepsilon _0) = 10^{ - 7} c^2$ & $\mu _0/(4\pi) = 10^{ - 7}$ & $1$ & $1$\\
\hline
\end{tabular}
\end{center}\vspace*{1ex}
where UES refers to the electrostatic cgs units system, UEM to the electromagnetic cgs system, H-L to the Heaviside-Lorentz system and the last line to the rationalized MKS system.  The first one is more appropriate to our intentions but, as we want to be closer to the Newtonian mechanics definitions,
we will use a units system in which electric charge is measured in units of mass, then only remain the mechanical units, the meter, the kilogram and the second. In fact the proposed system of units is simply MKS without adding any additional unit defined from the electromagnetic equations. To emphasize this fact we call this system, specially when is used in electromagnetism, pure MKS system of units:
\begin{center}
\begin{tabular}{| l | c | c | c | c |}
\hline
  &$ k_1 $& $ k_2 $& $ k_3 $& $ k_4 $\\
\hline
MKS-pure	& $G$ & $G c_{}^{ - 2}$ & $ 1 $ & $ 1 $\\ 
\hline
\end{tabular}
\end{center}
where $ G $ is the gravitational constant. In this pure MKS  system the Maxwell-Lorentz equations are
\begin{equation}\label{eq15}
m\frac{{d^2 \vec x}}
{{dt^2 }} = m_e \;\left( {\vec E + \;\frac{{d\vec x}}
{{dt}} \times \vec B} \right)
\end{equation}
\begin{equation}\label{eq16}
\left. \begin{gathered}
  \vec \nabla  \times \vec E =  - \frac{{\partial \vec B}}
{{\partial t}} ;\qquad \vec \nabla \cdot\vec B = 0 \hfill \\
  \vec \nabla \cdot\vec E = 4\pi G\rho _e ;\quad \vec \nabla  \times \vec B = \frac{1}
{{c^2 }}\left( {4\pi G\vec J_e  + \frac{{\partial \vec E}}
{{\partial t}}} \right) 
\end{gathered}  \right\}
\end{equation}
We have changed the charge symbol $ q $ to the more convenient $ m_e$ to emphasize the fact that in this units system masses and charges are measured in the same mass units.

As it is well know, the equation where the sources are not involved,  $\vec \nabla \cdot \vec B = 0 $, guarantees the existence of the vector potential $\vec A $ through $\vec B=\vec \nabla \times\vec A $. Now  the dimensions of $\vec B$ are $T^{-1}$ so $\vec A $ has the dimensions of a velocity. The remaining  equation without sources lead to 
$$\vec \nabla \times \left(\vec E + \frac{\partial \vec {A}}{\partial t}\right)=0$$
which tells us that $ \vec E +{\partial\vec{A}}/{\partial t}$ is the gradient of a potential $\phi$ whose dimensions are the square of a velocity. 
We can also see that in the limit $c\to \infty $ we get the same Lorentz equation and whereas  Maxwell equations become
\begin{equation}\label{eq17}
\left. \begin{gathered}
  \vec \nabla  \times \vec E =  - \frac{{\partial \vec B}}
{{\partial t}} ;\qquad \vec \nabla \cdot\vec B = 0 \hfill \\
  \vec \nabla \cdot\vec E = 4\pi G\rho _e ;\quad \vec \nabla  \times \vec B = 0 
\end{gathered}  \right\}
\end{equation}
Notice the similarity of these Maxwell and Lorentz equations when there is no charge density,  $ \rho_e = 0$, with the non-inertial fields equations \eqref{eq09}. It can be even closer if for the electromagnetic fields we use the gauge $\phi = (\vec A)^2/2$, then  we have \eqref{eq16} and
\begin{equation}\label{eq18}
\vec E=\vec \nabla \left( {\frac{{\vec{A}^2 }}
{2}} \right) -\frac{{\partial \vec A }}{{\partial t}};\quad\vec B  =   \vec \nabla  \times \vec A,
\end{equation}
which are almost the same as \eqref{eq08} where the vector potential plays the role of the velocity field in rigid motions.
But the Gauss equation  for electromagnetic field,  namely  that with $\nabla\cdot\vec E $, doesn't contain a term $\vec B^2/2 $ as it is contained in the counterpart equation for the velocity field \eqref{eq09} for rigid motions.

Now, in the usual conditions of null fields at infinity, equations \eqref{eq15} and \eqref{eq16} become 
\begin{equation}\label{eq18b}
m\vec a = m_e \;\vec E
\end{equation}
\begin{equation}\label{eq19}
\left. \begin{gathered}
  \vec \nabla  \times \vec E = 0 \hfill \\
  \vec \nabla \cdot\vec E = 4\pi G\rho _e \; \hfill \\ 
\end{gathered}  \right\}
\end{equation}
The magnetic field does not appear due to the fact that the equations for the magnetic field are $ \vec\nabla\cdot\vec B = 0,$ and $\vec\nabla\times\vec B = 0$ and the only solution vanishing at infinity is $\vec B=0$. A remarkable fact is that according to the limit $ c\to\infty$ of the Maxwell equations, written in the system of units that we propose, \textit{the magnetic field is entirely a relativistic effect}.

In this way the likeness between  electromagnetism and gravitation increases. The equations for the gravitational (non-relativistic) interaction  are
\begin{equation}\label{eq20}
m\vec a = m_g  \vec g ;\quad (m = m_g )
\end{equation}
\begin{equation}\label{eq21}
\left. \begin{gathered}
  \vec \nabla  \times \vec g = 0 \hfill \\
  \vec \nabla \cdot\vec g =  - 4\pi G\rho \; \hfill \\ 
\end{gathered}  \right\}
\end{equation}
To stress the differences between the electromagnetic non-relativistic equations and the gravitational Newtonian, notice that
\begin{center}
\begin{tabular}{| l | l | }
\hline
  Electromagnetism &	Gravitation\\
\hline
 $ m \ne m_e $ &  $ m = m_g $\\
\hline
  $ G $ &  $ -G $\\
\hline
\end{tabular}
\end{center}

\section{\label{Sec04} Generalized non-relativistic gravitation and equivalence principle}
So far, except for equation \eqref{eq08}, we have only obtained expressions that were already known and that perhaps we have written in a unusual way. We shall now attempt a revision of Newtonian theory of gravity, aiming to get a better understanding of General Relativity theory. To this purpose we start analyzing the weak relativistic equivalence principle:\\
\textbf{Relativistic equivalence principle}\\
\textbf{\textit{At every space-time point in an arbitrary gravitational field it is possible to choose a ``locally inertial system of coordinates'' such that, within a sufficiently small region around this point the laws of mechanics are the same as in an inertial Cartesian coordinate system in the absence of gravitation.}}

Notice that this principle says nothing about how to built the ``locally inertial coordinate system''. It only states its existence.

In section \ref{Sec03} when studying the forces of inertia for rigid motions, we have seen that a velocity field opened the possibility to define a local frame, and how the velocity field of the rigid motion $\vec{v}_0$ can be used as a vector potential to define the acceleration and the rotation fields. Also in section \ref{Sec03} we used  a gauge where only a vector potential, with dimensions of velocity, is needed to build the electromagnetic field.

In a similar way as we did with the Newtonian rigid velocity field, we shall introduce a generalized gravitational vector potential from which the
gravitational field can be derived.

Recall that from the viewpoint of the inertial frame there is no force field for a free particle but acording to a non-inertial frame ---which follows a Newtonian rigid motion--- two inertial force fields arise which can be associated to a vector potential,  $\vec v_0 (\vec x,t)$, (the rigid velocity field). 

Assume now that we have a Newtonian gravitational field in an inertial frame. We want to introduce a velocity field $\vec v_g (\vec x,t)$ such that a particle in the local non-inertial frames associated  to it does not feel any field of force.  

At each point in space and for every time we define the local system as a reference frame whose origin moves with the  velocity $\vec v_g (\vec x,t)$, i.e. the interaction vector potential, and with a triad of orthogonal axes (with respect to the Euclidean metric)  that ``rotates rigidly'' with a local angular velocity $\vec \Omega_g=\frac12 \nabla\times \vec v_g(\vec x,t)$, that is the vorticity of the velocity field.  Now our \textbf{non-relativistic equivalence principle} states that\\
\textbf{\textit{The trajectory of a particle due to gravitational interaction $\vec v_g (\vec x,t)$, is such that with respect to the origin of the local system has no acceleration.}}

Notice that unlike  the weak relativistic equivalence principle in this non-relativistic equivalence principle we know the motion of the local frame.

We know that for a velocity field $\vec v_g(\vec x,t)$ the acceleration field is given by
\begin{equation}\label{eq22}
\vec{A}(\vec x,t)=\frac{{\partial \vec v_g }}
{{\partial t}}+(\vec v_g \cdot \vec \nabla)  \vec v_g =\frac12\nabla(\vec v_g)^2+\frac{{\partial \vec v_g }}{{\partial t}}+2\vec \Omega_g\times \vec v_g
 \end{equation}
and this will be the acceleration of our local frame.

The relation between the accelerations referred to the inertial system and to the local one is (see Annex I)\eqref{eqA3}:
$$\vec a=\vec b+\vec A+\frac{d\vec\Omega}{dt}\times \vec y + \vec\Omega\times(\vec\Omega\times\vec y)+2\vec\Omega\times\vec w$$
Consider now a particle at $\vec x$ (in the inertial frame) in a gravitational field. As seen from the local reference it is at the origin, $\vec y=0$, without acceleration and with velocity $\vec w$. By the non-relativistic equivalence principle, its acceleration referred to the inertial frame is: 
\begin{equation}\label{eq23}
\vec a = \frac12\nabla(\vec v_g)^2+\frac{{\partial \vec v_g }}{{\partial t}}+2\vec \Omega_g\times \vec v,
\end{equation}
where we have included that $\vec v_g+\vec w$ is the particle velocity in the inertial system $\vec v$.
The acceleration produced by our generalized Newtonian gravitational fields has two components, one depending on the particle velocity (as the Coriolis component of non-inertial fields) and the other that does not depend on the particle velocity
\begin{equation}\label{eq24}
\vec a =  \vec g +  \vec v \times \vec \beta 
\end{equation}
This equation is very similar to the Lorentz equation, the only difference is that the gravitational charge coincides with the inertial mass, a feature which is on the basis of the non-relativistic equivalence principle. We have introduced the acceleration and rotation gravitational fields as follow:
\begin{equation}\label{eq25}
\vec g \equiv \vec \nabla \left( {\frac{{\vec v_g^2 }}
{2}} \right) + \frac{{\partial \vec v_g }}
{{\partial t}} ;\quad \vec \beta  \equiv  -\vec \nabla  \times \vec v_g 
\end{equation}
Notice that with this definition $\vec\beta= -2\vec\Omega_g$. Furthermore we have the field equations
\begin{equation}\label{eq26}
\vec \nabla  \times \vec g =  - \frac{{\partial \vec \beta }}
{{\partial t}};\quad \vec \nabla \cdot\vec \beta  = 0
\end{equation}
which are equivalent to the first pair of Maxwell equations.
Now we can add the source equation for non-vanishing mass density
\begin{equation}\label{eq27}
\vec \nabla \cdot\vec g =  - 4\pi G\rho 
\end{equation}
and a non-relativistic version of the Ampere law 
\begin{equation}\label{eq28}
\vec \nabla  \times \vec \beta  = 0
\end{equation}
so we have a system of gravitomagnetic equations like the non-relativistic Maxwell equations \eqref{eq17}.

The condition that the fields vanish at infinity implies $\vec\beta  = 0$ and we obtain the usual results, the equation of motion is
$\vec a =  \vec g; $
and the source equation \eqref{eq27}.  

We can go further and propose a gravitational interaction that includes the inertial fields. We keep the equation of motion \eqref{eq24} and propose the following field equations 
\begin{equation}\label{eq30}
\left. \begin{gathered}
  \vec \nabla  \times \vec g =  - \frac{{\partial \vec \beta }}
{{\partial t}} ;\quad \vec \nabla \cdot\vec \beta  = 0 \hfill \\
  \vec \nabla \cdot\vec g - \frac{1}
{2}\;\;\vec\beta _{}^2  =  - 4\pi G\rho ;\quad \vec \nabla  \times \vec \beta  = 0 \hfill \\ 
\end{gathered}  \right\}
\end{equation}

Equations \eqref{eq30} are the equations for the gravitational field as seen by any observer, inertial or not. If we don't have mass density, the velocity field is the non-inertial one. If we don't have fields at infinity the solutions are the usual ones in Newtonian gravitation. But if there is a mass density and fields do not vanish at infinity we have something new.

An interesting fact is that the ``magnetic gravitational field''   $ \vec \beta $ is a non-relativistic effect. This is truly remarkable because in section \ref{Sec03} we have arrived to the conclusion that the  magnetic  field $ \vec B $ in electrodynamics is a purely relativistic effect. 

\section{\label{Sec05}Lagrangian and Hamiltonian for the non-relativistic generalized gravitational interaction}

Given the field of  gravitational interaction $\vec v_g (\vec x,t)$, from the equivalence principle in the non-inertial reference system the particle is free. So in the Lagrangian only the kinetic term appears $ L=m w^2/2$ which can be written in terms of the velocity in the inertial system
\begin{equation}\label{NL}
L = \frac{1}
{2}m(\dot{\vec x} - \vec v_g(\vec x,t) )^2 
\end{equation}
The Euler-Lagrange equations are:
\begin{eqnarray}\label{EL}
\frac{d\;}{dt}\frac{\partial L}{\partial \dot{\vec x}}-\frac{\partial L}{\partial \vec x}&=&m\left(\ddot{\vec x} -(\dot{\vec x}\cdot \nabla)\vec v_g(x,t)-\frac{\partial \vec v_g}{\partial t}\right)\nonumber \\
&&\hspace*{-1em}-m\left(\frac12\nabla \vec v^2_g-\nabla(\dot{\vec x}\cdot\vec v_g)\right)=0
\end{eqnarray}
Now using the identity
$$\nabla(\vec A\cdot\vec B)=\vec A \times (\nabla \times \vec B)+\vec B \times (\nabla \times \vec A)+(\vec A\cdot \nabla)\vec B+(\vec B\cdot \nabla)\vec A$$
and taking into account that $\dot{\vec x}$ and $\vec x$ are independent variables (in phase space), we have that
$$\nabla(\dot{\vec x}\cdot\vec v_g)-(\dot{\vec x}\cdot \nabla)\vec v_g(x,t)=\dot{\vec x} \times (\nabla \times \vec v_g(\vec x,t))$$
which substituted in \eqref{EL} leads to the  equations of motion \eqref {eq23}, \eqref {eq24}. 
We can also also set up the Hamiltonian formulation. First performing the Lagrange transformation
$$\vec p =\frac{\partial L}{\partial \dot{\vec x}}=m(\dot{\vec x} -\vec v_g(\vec x,t) )$$
then the energy function is
\begin{equation}\label{eq32}
{\cal E} = \vec p\cdot \dot{\vec x}-L=\frac{1}{2}m\dot{\vec x}^2  - \frac{1}{2}m\vec v_g^2 
\end{equation}
and the Hamiltonian
\begin{equation}\label{eq33}
H = \frac{{\vec p}\,^2 }{2m} + \vec p\cdot\vec v_g 
\end{equation}

\section{\label{Sec06}Non-relativistic limit of Einstein's equations}
The free particle relativistic Lagrangian related to \eqref{NL} is
\begin{equation}\label{eq34}
L =  - mc^2\sqrt{1-\frac{(\dot{\vec x} - \vec v_g )^2}{c^2}} 
\end{equation}
and the action functional
$$ S=-mc\int ds =\int Ldt= -mc\int\sqrt{c^2-(\dot{\vec x} - \vec v_g )^2}dt$$
this can be associated to a metric
\begin{equation}\label{eq35}
\begin{gathered}
ds^2=-\left({c^2-(\dot{\vec x} - \vec v_g )^2 }\right)dt^2 =\\ -(c^2 - \vec v_g^2 )dt^2+d\vec x^2-2\vec v_g \cdot d\vec x  dt
\end{gathered}
\end{equation}

which has the following properties:

1) This is a so-called Newtonian metric \cite{LB} invariant under the rigid motion group. The transformation of the metric tensor under a change of coordinates associated to the rigid motion $x^i=X(t)^i+R^i_j y^j$ and  $t=t'$, is
$$g_{0'0'}=g_{00}+2 A^s_{0'} g_{0s}+A^n_{0'}A^m_{0'} g_{nm}$$
$$g_{j'0'}={R}^i_{j'}g_{0i}+R^n_{j'}A^m_{0'}g_{nm}$$
$$g_{i'j'}={R}^n_{i'}{R}^m_{j'}g_{nm}$$
where $$A^s_{0'}=\dot{R}^s_{j'}y^{j'}+\dot X^s$$
That is, the metric given by the expression \eqref{eq35} is of this type in every non-inertial frame. More specifically, the metric is invariant provided that the velocity field transforms as $\vec{v}_g\rightarrow \vec{v}'_g=\vec{v}_g-(\vec{\Omega}\times\vec y+\dot{\vec X})$ 

2) In the limit $ c \to \infty $ the metric connection for \eqref{eq35}  leads to the Newtonian affine connection, because the inverse metric in the $ c \to \infty $ limit is  $ g^{0\mu }  = 0\;\;\;g^{ij}  = \delta ^{ij}$
and neglecting the small terms when $ c \to \infty $ the Christoffel symbols are
\begin{equation}\label{eq36}
\begin{gathered}
\Gamma _{jk}^i  = 0 ;\quad \Gamma _{\mu \nu }^0  = 0 ;\quad \Gamma _{00}^i  = \partial _i \left( {\frac{{v_g^2 }}
{2}} \right) + \partial _t v_g^i  \\
\Gamma _{j0}^i  = \frac{1}{2}(\partial _j v_g^i  - \partial _i v_g^j )
\end{gathered} 
\end{equation}
3) The Ricci tensor of the metric \eqref{eq35} when $ c \to \infty $ is:
\begin{equation}
R_{ij}  = 0;\quad R_{00}  = -\left(\vec \nabla \cdot\vec g  - \frac{\beta^2}{2}\right)
;\quad R_{0i}  =\frac12 (\vec \nabla  \times \vec \beta )_i^{}  
\end{equation}
This means that, using the metric \eqref{eq35} the  Einstein's equations for vacuum, $ R_{\mu \nu }  = 0$ in the limit  $ c \to \infty $  yield the  Newtonian equations \eqref{eq30} for $\rho=0$. It is very interesting to note that, without any other consideration, the limit $ c \to \infty $ yields the right Newtonian limit.

4) Another interesting result, showing the strength of this way to obtain a relativistic mechanics from this Newtonian generalization, is that the velocity field  needed to obtain the spherical symmetric solutions for vacuum in the Newtonian and in the relativistic case are the same. 
If we take
$$\vec v_g=v_g(r) \hat r$$ where $\hat r$ is the unit radial unitary vector, then the vacuum solution of the metric \eqref{eq35} is the Schwarzschild solution and $v_g(r)$ function must be 
\begin{equation}\label{eq37} 
v_g(r)=\sqrt{\frac{k}{r}}
\end{equation}
but what is not usually done is to write it in this form \cite{Painleve}. 

The same function \eqref{eq37} using  \eqref{eq25} gives the solution for the Newtonian equations \eqref{eq30} with $\rho=0$.

\section{Conclusions} 
We study the velocity field for the classical rigid motion and derive the field equations for its acceleration and rotation fields, the inertial fields. We also write the Maxwell-Lorentz equations for the electromagnetic field using a suitable system of units and  a gauge where only the  potential vector, which has the dimensions of a velocity, is necessary. For them we find the limit for $c\to\infty$. Taking into account the null limit condition at infinity of the electromagnetic field we get the same equations, with some change in the constants, as for the Newtonian gravitational problem. 

We build a generalized Newtonian gravitational theory where the potential is a velocity field as in the two previous examples and that includes in an unified way the inertial forces fields and the gravitational Newtonian field. If the fields are nulls at infinity only the gravitational field remain, and if the mass density is zero we obtain the equations for inertial fields. We introduce a non-relativistic equivalence principle which is very useful to construct a Lagrangian theory for test particles moving in this generalized Newtonian gravitational field.
Finally we use this Lagrangian to build a relativistic theory where we have a Newtonian metric invariant for the rigid motion group. Some interesting features of this metric are that the limit metric connection when  $c\to\infty$ is the Newtonian affine connection, and the Einstein's equations in the same limit lead to the Newtonian field equations wihtout using any kind of weak field aproximation.

Another interesting fact is that the same spherical symmetric velocity field $\sqrt(k/r)\hat r$ gives us the Schwarzschild solution when we put it in the metric and the Newtonian known result when we use the Newtonian field equations.

\section*{ANNEX I} 
We give here a detailed derivation of the transformation of position, velocity and acceleration between two orthonormal reference frames.
We first change to a non-inertial reference system moving rigidly. From the Chasles theorem the most general motion of this frame is such that its origin moves arbitrarily with respect to the origin of an inertial frame and its axis rotates with angular velocity $\vec \Omega(t)$.

We have an inertial frame with origin in  $O(0,0,0)$ and a system of orthonormal axis $\vec\varepsilon_i,\; i=1\ldots 3$ and one non-inertial frame with origin  $Q(X^1(t),X^2(t),X^3(t))$ and three orthonormal axis $\vec e_i(t),\; i=1\ldots 3$
$$
\vec\varepsilon_i\,\vec\varepsilon_j = \vec e_i(t)\,\vec e_j(t) = \delta_{ij}\ \forall t
$$
$$
\vec e_j(t) = R^i_j(t)\,\vec\varepsilon_i 
$$
where $R^i_j(t)$ is a rotation matrix. A point $P$ can be referred to both frames, as $\vec x$ in the inertial frame and as $\vec y$ in the non-inertial frame
$$
\vec x = \vec X + \vec y ;\quad \big(x^i\vec\varepsilon_i =X^i\vec\varepsilon_i +y^j\vec e_j\big)$$
or
\begin{equation}\label{eqA1}
x^i = X^i(t) + R^i_j(t)\,y^j
\end{equation}
The velocity transformation is:
$$
\vec v \equiv \frac{dx^i}{dt}\vec \varepsilon_i = \frac{dX^i}{dt}\vec\varepsilon_i +
\frac{dy^j}{dt}\,\vec e_j + y^j\frac{d\vec e_j}{dt} 
= \vec V + \vec w + y^j\frac{d\vec e_j}{dt} 
$$
where
$$
\frac{d\vec e_j}{dt} =\dot R^k_j\,\vec\varepsilon_k = \vec\Omega\times\vec e_j,$$
and
$$\vec\Omega \equiv \frac12\sum_j R_j^k\,\dot R_j^l\,
\vec\varepsilon_k\times\vec\varepsilon_l = \Omega^h \vec\varepsilon_h
$$
$$
\Omega_h \equiv \frac12 \sum_jR_j^k\,\dot R_j^l\eta_{klh} ;\quad \Omega^{kl} =\sum_j R_j^k\,\dot R_j^l.
$$
So it can be also written as
\begin{equation}\label{eqA2}
 \vec v = \vec w + \vec V + \vec\Omega\times\vec y \quad 
\end{equation}
or alternatively
$$
\dot x^k =  \dot X^k + R^k_l \dot y^l + \dot R^k_l y^l \quad 
$$

\noindent Note
$$
\vec\Omega\times\vec y = \frac12 \sum_j R_j^k\,\dot R_j^l\,
(\vec\varepsilon_k\times\vec\varepsilon_l)\times(y^i R_i^h \vec\varepsilon_h) = \dot R_j^k y^j \vec\varepsilon_k
$$

\noindent The acceleration transformation is
$$
\vec a \equiv \frac{d^2x^i}{dt^2}\vec\varepsilon_i = \frac{d^2y^i}{dt^2}\vec e_i +
\frac{dy^i}{dt}\frac{d\vec e_i}{dt} + \frac{d^2X^i}{dt^2}\vec\varepsilon_i+ $$
$$\frac{d\vec\Omega}{dt}\times\vec y +\vec\Omega\times\left(\frac{dy^i}{dt}\vec e_i + y^i\frac{d\vec e_i}{dt}\right)
$$
That can be also written
\begin{equation}\label{eqA3}
\vec a = \vec b + \vec A + \,\frac{d\vec\Omega}{dt}\times\vec y +
\vec\Omega\times(\vec\Omega\times\vec y) + 
2\vec\Omega\times\vec w
\end{equation}
or
$$
\ddot x^k = R^k_l \ddot y^l + \ddot X^k +\ddot R^k_l y^l+  2\dot R^k_l\dot y^l $$

\noindent Let's remark
$$
\frac{d\vec\Omega}{dt}\times\vec y+\vec\Omega\times(\vec\Omega\times\vec y) = \ddot R^k_l y^l\vec \varepsilon_k ;\quad \vec\Omega\times\vec w = \dot R_j^k \dot y^j \vec \varepsilon_k 
$$

\begin{acknowledgments}
Thanks to three good friends,   to Jes\'{u}s Mart\'{\i}n of who we have copied the Annex, to Josep Llosa for doing a carefully reading and an useful criticism that lead us to improve this paper and finally to Llu\'{\i}s Bel that without his inspiration and insistence almost nothing of this article had occurred to us.
\end{acknowledgments}

\end{document}